\documentclass[conference, 10pt]{IEEEtran}
\usepackage{amsmath,amsfonts}
\usepackage{algorithm}
\usepackage[noend]{algpseudocode}
\usepackage{array}
\usepackage[caption=false,font=normalsize,labelfont=sf,textfont=sf]{subfig}
\usepackage{textcomp}
\usepackage{stfloats}
\usepackage{url}
\usepackage{verbatim}
\usepackage{graphicx}

\usepackage{amssymb}
\usepackage{amsthm}
\usepackage{bm}
\usepackage{booktabs}
\usepackage{color}
\usepackage{cite}             
\usepackage{comment}          
\usepackage{enumitem}
\usepackage{etoolbox}         
\usepackage{graphicx}         
\usepackage{mathtools}
\usepackage{multirow}
\usepackage{pgfplots}
\usepackage{pgfplotstable}
\usepackage{setspace}
\usepackage{stfloats}
\usepackage{caption}
\usepackage{subfig}
\usepackage{tabularx}
\usepackage{threeparttable}
\usepackage{tikz}
\usepackage{url}
\usepackage{xcolor}
\usepackage{balance}
\usepackage[table]{xcolor}

\pgfplotsset{compat=newest}

\newcolumntype{Y}{>{\centering\arraybackslash}X}

\theoremstyle{definition}

\renewrobustcmd{\bfseries}{\fontseries{b}\selectfont}
\renewrobustcmd{\boldmath}{}
\newrobustcmd{\B}{\bfseries}

\usetikzlibrary{decorations.pathreplacing,calc}

\usepackage{soul}
\definecolor{xiaohan}{RGB}{139, 0, 18}

\usepackage{eso-pic}
\AddToShipoutPictureBG*{\AtPageUpperLeft{\put(0,-14){\makebox[\paperwidth][c]{\parbox[t]{0.86\paperwidth}{\centering\fontsize{7}{8}\selectfont
\copyright~2026 IEEE. Personal use of this material is permitted. Permission from IEEE must be obtained for all other uses, in any current or future media, including reprinting/republishing this material for advertising or promotional purposes, creating new collective works, for resale or redistribution to servers or lists, or reuse of any copyrighted component of this work in other works.
This is the accepted version of the paper published in the 2026 IEEE Symposium on Field-Programmable Custom Computing Machines (FCCM), DOI: 10.1109/FCCM68464.2026.00029.\\[1pt]
\textbf{arXiv v2: this version corrects the total-wirelength (tot WL) results in Table~II. See the Correction Notice at the end of the paper.}}}}}}

\graphicspath{./figs/}

\definecolor{xiaohan}{RGB}{139, 0, 18}

\usepackage{threeparttable}
\usepackage{graphicx}
\usepackage{subcaption}

\begin{document}

\title{
DiffRouter: A Differentiable Routing Framework \\ for UltraScale FPGAs
}

\author{
    \IEEEauthorblockN{Xiaohan Gao$^{1}$, Zhili Xiong$^{1}$, Yusu Wang$^{2}$, and David Z. Pan$^{1}$}
    \IEEEauthorblockA{$^{1}$Department of Electrical \& Computer Engineering, The University of Texas at Austin, TX, USA \\
    $^{2}$Halıcıoğlu Data Science Institute, University of California, San Diego, CA, USA\\
    \{xiaohan.gao@austin., zhilix691\}@utexas.edu, yusuwang@ucsd.edu, dpan@ece.utexas.edu}
}

\maketitle

\begin{abstract}

Routing has become a major bottleneck in large-scale FPGA design flows, where traditional routers rely on sequential net-by-net processing and rip-up--reroute heuristics. 
This inherent sequentialism limits parallelism, leads to suboptimal utilization of routing resources, and significantly increases runtime on modern FPGAs.
This paper introduces \textit{DiffRouter}, the first differentiable, gradient-based FPGA routing framework with GPU acceleration. 
By lifting routing into a continuous optimization space, \textit{DiffRouter} enables highly concurrent routing and improved global optimality before projecting solutions back onto the discrete fabric. 
The framework consists of three key stages:
(1) \textbf{Resource-pruned preprocessing}, which extracts compact routing variables per net to reduce memory footprint and problem complexity;
(2) \textbf{Differentiable global routing via gradient descent}, which formulates routing as a Lagrangian-relaxed optimization that minimizes wirelength while iteratively enforcing connectivity and congestion constraints as relaxed terms; and
(3) \textbf{Postprocessing}, which maps the continuous wire distributions to a valid UltraScale routing solution and completes solutions by detailed routing.
Experiments on the FPGA 2024 Routing Contest benchmarks show that \textit{DiffRouter} outperforms state-of-the-art parallel routers, achieving over \(5\times\) speedup over RWRoute and \(2\times\) over Vivado, and running \(18\%\) faster than the state-of-the-art academic FPGA router Potter while delivering competitive critical path wirelength.
\end{abstract}

\section{Introduction} 
\begin{figure}[tb]
    \centering
                \includegraphics[width=0.92\columnwidth]{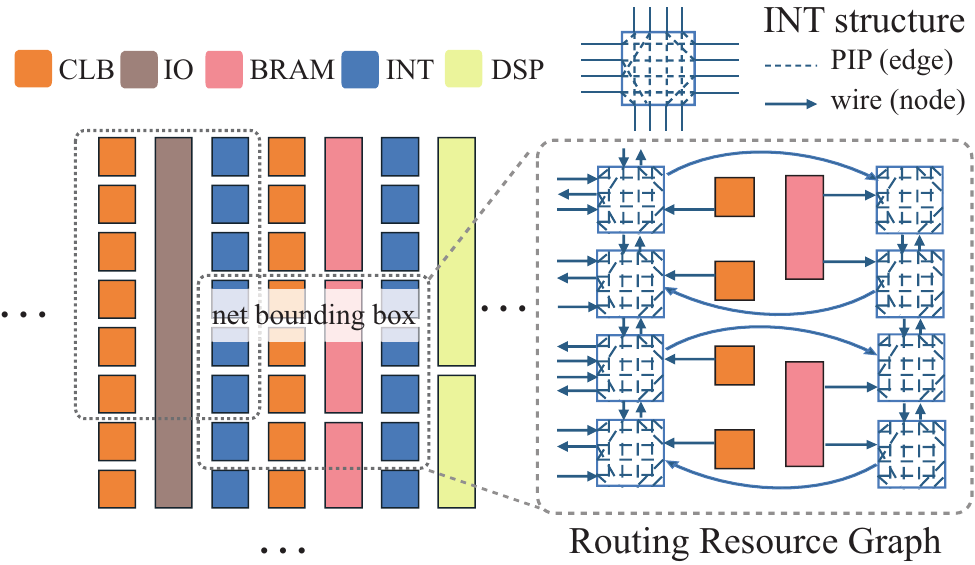}
    \caption{Routing resources of AMD-Xilinx Ultrascale+ FPGA. The FPGA canvas was divided into columns of different resource types, while the routing resources span across INT tiles (also known as switch boxes). Each INT tile connects to CLB or DSP/BRAM tiles on its left and right sides. 
    }
    \label{fig:rrg}
\end{figure}

Routing is a critical stage in the FPGA physical implementation flow, following placement, which assigns exact locations to all logical components on the FPGA fabric.
Unlike ASIC routing, where the solution space is larger and more flexible, FPGA routing must operate within fixed, highly constrained routing resources that can be represented as a routing resource graph (RRG), as illustrated in Fig.~\ref{fig:rrg}.
The objective is to map each wire segments onto this graph through programmable interconnect points (PIPs), minimizing total wirelength and critical-path length while preserving connectivity and avoiding capacity violations. 
Achieving these goals often requires many rip-up and re-route iterations, leading to prolonged router runtime.

Prior academic work has explored various approaches to accelerate FPGA routing.
Most techniques exploit net-level parallelism and scheduling~\cite{ParaDRo, ParRA, recurPart, FPGA_TRETS2020_Zhou, FPGA_FPL2024_Martin, FPGA_ICCAD2025_Zang}, demonstrating substantial runtime reduction.
These methods recursively partition the netlist based on the spatial distribution of net bounding boxes, enabling each partition to be routed concurrently under the assumption of minimal inter-partition routing dependencies.
Within each thread, a negotiated-congestion router PathFinder~\cite{PathFinder} is used to resolve conflicts and converge to a legal solution.
A more fine-grained single-net-level parallelism was proposed in~\cite{Corolla}, where a GPU-friendly Bellman–Ford kernel serves as a fast single-source shortest-path solver within PathFinder.
The VPR tool~\cite{VPR8} further improves sequential PathFinder by introducing incremental routing, which prunes illegal routing subtrees to reduce rip-up overhead.
\label{sec:intro}
\begin{figure}[!t]
    \centering
    \includegraphics[width=1\columnwidth]{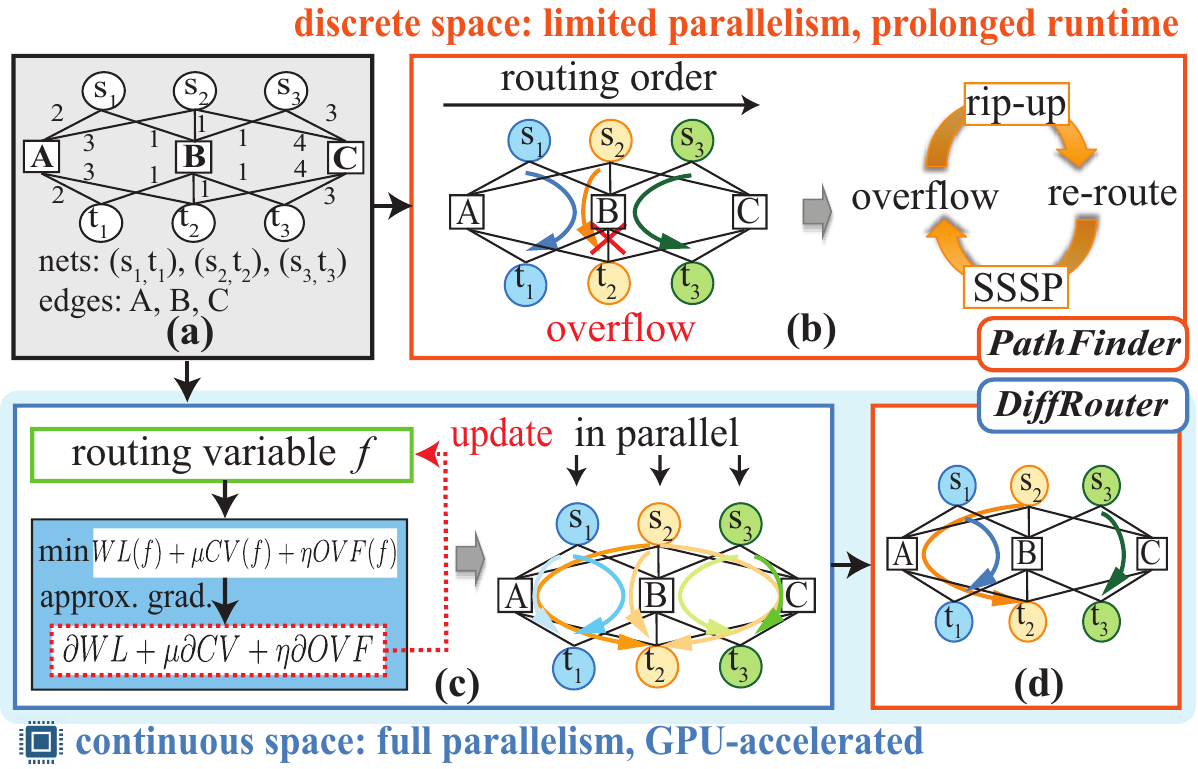}
    \caption{(a) Input nets on the RRG with node wirelength costs; (b) PathFinder flow in discrete space; (c) \textit{DiffRouter} global routing formulation in continuous space; (d) \textit{DiffRouter} detailed routing from continuous variables to discrete routes}
    \label{fig:motivation}
    \vspace{-0.2in}
\end{figure}
Despite these advances, existing approaches still rely heavily on heuristics, including sequential route net ordering, overlap tolerance in partitioning, and congestion factor tuning, etc., largely due to the discrete and cost-function-centric nature of the PathFinder algorithm.
As design sizes continue to grow, these heuristic constraints limit scalability and optimality, making it increasingly difficult for traditional FPGA routers to approach higher quality routing solutions.
Although recent machine learning based routing methods~\cite{GlobalRoutingOrdering, DeepRoute, RLripper} aim to learn optimal net ordering, select nets for rip-up and reroute, or prune routing resources, they often suffer from limited generalizability across architectures and designs.
This motivates our exploration of a global, fully differentiable formulation for FPGA routing as demonstrated in Fig.~\ref{fig:motivation}. 
Inspired by analytical placers, we construct a differentiable wirelength objective with relaxed connectivity and congestion term and optimize it through gradient descent, enabling substantially larger search space and higher parallelism. 
The resulting continuous solution is then legalized onto the discrete routing resource graph.

To validate our idea, we leverage the 2024 Runtime-First FPGA Interchange Routing Contest~\cite{ISFPGA_Routing_Contest}, which promotes both runtime speedup and critical-path wirelength (CPWL) reduction, as well as adoption of the FPGA Interchange Format (IF). 
In this work, we propose \textit{DiffRouter}, the first FPGA router to simultaneously employ a differentiable formulation and GPU acceleration for FPGA global routing. 
\textit{DiffRouter} achieves results that surpass state-of-the-art FPGA routers on the contest benchmarks. 
Our main contributions are as follows:
\begin{itemize}
    \item \textit{DiffRouter} is the first FPGA router to jointly leverage a fully differentiable formulation and GPU acceleration: it preprocesses and prunes routing variables, optimizes routing demand in continuous space, and then legalizes the solution onto the discrete FPGA routing resource.
    \item We formulate relaxed routing as a unified analytical objective consisting of a linear wirelength term, a connectivity term inspired by a spectral approach of the shortest path problem, and a congestion term based on relaxed network overflow.
    \item To enable multi-objective routing optimization with gradient-based solvers, we propose a novel approximation of connectivity gradients with faster matrix operations.
    \item On the 2024 Runtime-First FPGA Interchange Routing Contest benchmarks~\cite{ISFPGA_Routing_Contest}, \textit{DiffRouter} outperforms the best academic baseline, Potter~\cite{FPGA_ICCAD2025_Zang}. 
    It is also approximately \(2\times\) faster than Vivado 2023.2 and \(5\times\) faster than RWRoute. 
\end{itemize}

In Section~\ref{sec:prelim}, we review the FPGA architecture and routing problem, and summarize prior FPGA routing approaches. 
Section~\ref{sec:alg} presents the algorithmic details of the proposed \textit{DiffRouter} pipeline. 
Section~\ref{sec:exp} reports results on the contest benchmark suite, together with two ablation studies that validate key design choices.

\section{Preliminaries} 
\label{sec:prelim}

In this section, we introduce the FPGA routing problem using the target AMD-Xilinx Ultrascale+~\cite{Xilinx_UltraScalePlus} device.

\subsection{FPGA Architecture and Routing Problem}

As illustrated in Fig.~\ref{fig:rrg}, the device view of an AMD-Xilinx UltraScale+ part~\cite{Xilinx_UltraScalePlus} features a column-based architecture in which instances are mapped onto different resource types, such as configurable logic blocks (CLBs), digital signal processing units (DSPs), block RAMs (BRAMs), and IO blocks. 
The \textit{Interconnect} (INT) tiles span across these resource columns and provide the routing resources.
Each INT tile contains thousands of programmable interconnect points (PIPs), with each PIP connecting a pair of wire segments that may span vertically or horizontally to other INT tiles at distances of 1, 2, 4, or 12 tiles.
Other tiles connect to adjacent INT tiles with wires to access the routing resources.

The \textit{Routing Resource Graph} (RRG) of an FPGA can be abstracted as a directed graph \(G(V, E)\), where nodes \(V\) represent pins and wire segments, and edges \(E\) represent PIPs, each of which provides a routing capacity of 1.
The FPGA routing problem aims to find paths on \(G\) that connect the source pin to all sink pins of each net. 
For a net \(n\) with a source pin \(p_{s_n}\) and a set of sink pins \(P_{t_n}\), all of these pins correspond to some nodes in \(V\). 
To establish the required logical connectivity, the objective is to identify a sub-tree from \(G\) that contains \(p_{s_n}\) and all \(P_{t_n}\) while providing valid connecting paths among them. 
Routing trees of different nets must remain disconnected; otherwise, short-circuit violations occur.

\subsection{PathFinder and Heuristic Routing Frameworks}
Conventional FPGA routers are largely based on PathFinder-style negotiated-congestion routing~\cite{PathFinder,zha2022revisiting}. 
As shown in Fig.~\ref{fig:motivation} (b), PathFinder iteratively routes nets using a shortest-path solver, allowing temporary overuse and resolving conflicts by dynamically updating per-node or per-edge costs based on both current and historical congestion.
Nets that traverse congested regions are penalized in subsequent iterations, which drives rip-up and re-route until all capacity violations are eliminated.

A number of extensions improve the efficiency and feature support of PathFinder-based routing. 
CRoute~\cite{vercruyce2019croute} introduces connection-based routing and augments the negotiation mechanism using a sharing factor for resources shared by multiple connections. 
AIR~\cite{aspdac2020_AIR} proposes incremental rerouting for high-fanout nets and an architecture-aware lookahead strategy, and has been integrated into VTR 8~\cite{VPR8}. 
RWRoute~\cite{FPGA_TRETS2021_Zhou} further supports practical routing features such as partial routing and timing modeling.

\subsection{Parallel and Differentiable FPGA Routing}
To improve runtime, recent frameworks exploit parallelism at different granularities~\cite{Corolla0,Corolla,FPGA_TRETS2020_Zhou,ParaFRo,ParaDRo,fpl2021shen,ParRA,FPGA_GLSVLSI2024_Zang,TDR,parcad2012gort,recurPart,moctar2018deterministic,icfpt2024kosar}. 
Several CPU-based approaches increase parallelism by partitioning nets into conflict-free groups and routing groups concurrently~\cite{TDR,ParaDRo,ParRA,fpl2021shen}, while others apply spatial partitioning of the routing fabric to reduce cross-region dependencies~\cite{parcad2012gort,recurPart}. 
Follow-up work further addresses determinism, load balance, and scalability challenges in spatial partitioning~\cite{ParaDRo,fpl2021shen,ParaFRo}. 
More fine-grained parallelism has also been explored by decomposing high-fanout nets into smaller components that can be processed concurrently~\cite{icfpt2024kosar}. 
On GPUs, Corolla accelerates routing by using GPU-friendly single-source shortest-path primitives within a PathFinder-style framework~\cite{Corolla0,Corolla}.
In addition, \cite{iccad2007roy} proposes a discrete Lagrange multiplier formulation combined with A* routing, which introduces continuous relaxation and optimization ideas to FPGA routing.
\cite{aspdac2023wang, tcad2025jiang} explore the idea of decoupling FPGA routing into global routing and detailed routing stages.

\section{Algorithm} 
\label{sec:alg}

In this section, we describe the overall framework and key algorithms in \textit{DiffRouter}. 
\subsection{Overview of the Framework}

\begin{figure}[tb]
    \centering
    \includegraphics[width=0.81\columnwidth]{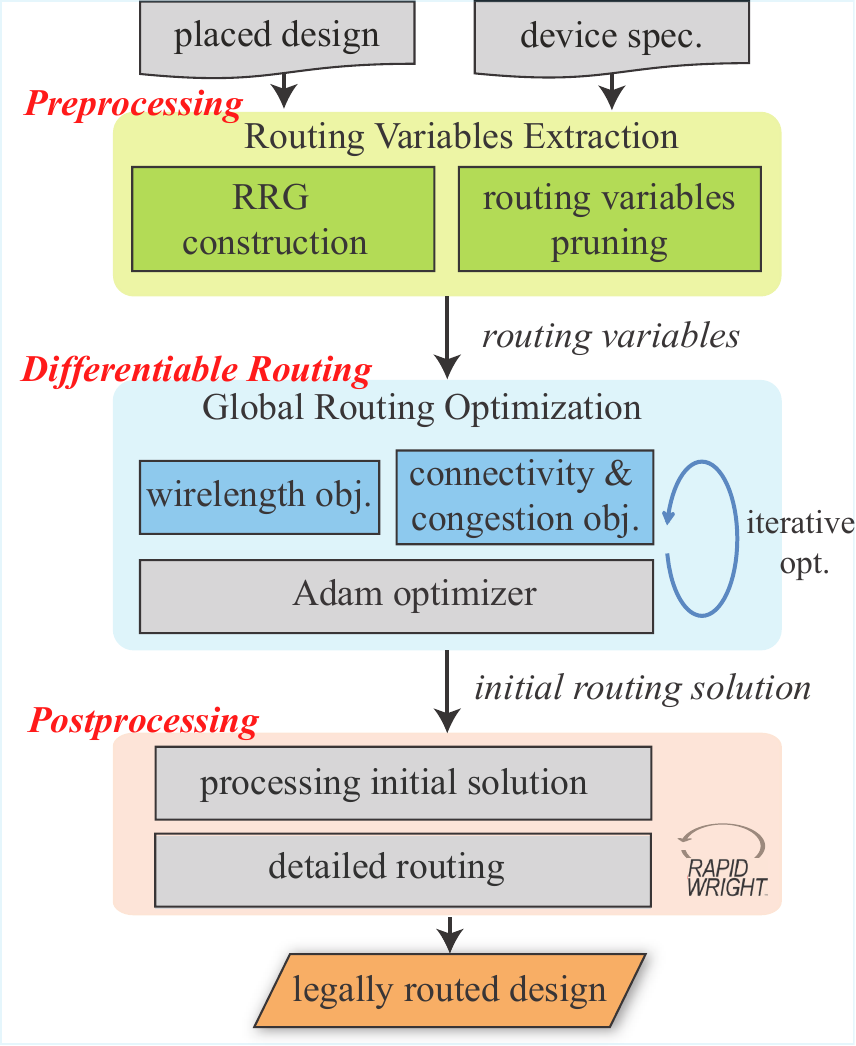}
    \vspace{0.1in}
    \caption{The overall flow of the proposed \textit{DiffRouter}}
    \label{fig:framework}
\end{figure}

Fig.~\ref{fig:framework} illustrates the overall flow of our routing framework \textit{DiffRouter}.
Our framework takes the physical netlist with placement information and FPGA device specification as input.
In order to achieve differentiable routing optimization, \textit{DiffRouter} abstracts routing resources as a graph representation.
The preprocessing stage defines the routing variables and constructs a routing resource graph (RRG) for the subsequent routing stage.

The differentiable routing stage addresses the concurrent routing of selected nets.
The core idea is to formulate the main objectives of routing as differentiable functions dependent on predefined routing variables.
Wirelength is the primary optimization objective.
Additional hard constraints used to ensure routing legality are relaxed into objective functions during the routing stage.
In particular, \textit{DiffRouter} considers path connectivity and routing congestion that guide the construction of valid routing solutions.

Different from the A-star-based routing methods, \textit{DiffRouter} utilizes the optimization process to produce routing solutions.
The optimization process generates global routing solutions, and the post-processing stage finalizes the routing results with detailed routing by connecting routing paths and re-routing illegal nets.

\begin{table}[tb]
    \centering
    \caption{Summary of Notation}
    \label{tab:notation}
    \resizebox{0.92\columnwidth}{!}{
    \begin{tabular}{ll}
        \toprule
        \textbf{Symbol} & \textbf{Definition} \\
        \midrule
        $G = (V, E)$        & Routing resource graph \\
        $n$                 & A net \\
        $p$, $P_{n}$        & A pin; set of pins of net $n$ \\
        $v$                 & A node in the routing graph \\
        $e$                 & An edge (PIP) in the routing graph \\
        $E^{-}(v), E^{+}(v)$ & Incoming and outgoing edges of node $v$ \\
        $w_e$               & Length (cost) of edge $e$ \\
        $f_e$               & Routing flow/variable on edge $e$ \\
        $\alpha_n$          & Importance/weight of net $n$ \\
        \bottomrule
    \end{tabular}
    }
\end{table}

\subsection{Routing Variables Extraction}
\label{sec:preprocessing}

The routing resources of FPGAs are highly regular, approximating a grid-like graph structure.
To enable the concurrent routing of massive nets, we adopt a simplified abstraction of the RRG for global routing. 

We define the simplified RRG as $G=(V, E)$. 
In $G$, each node $v\in V$ represents an INT tile, and each edge $e$ represent inter-tile connection with capacity $c$.
The capacity $c$ is estimated by the number of physically feasible wires.

\begin{figure}[tb]
    \centering
    \includegraphics[width=0.99\columnwidth]{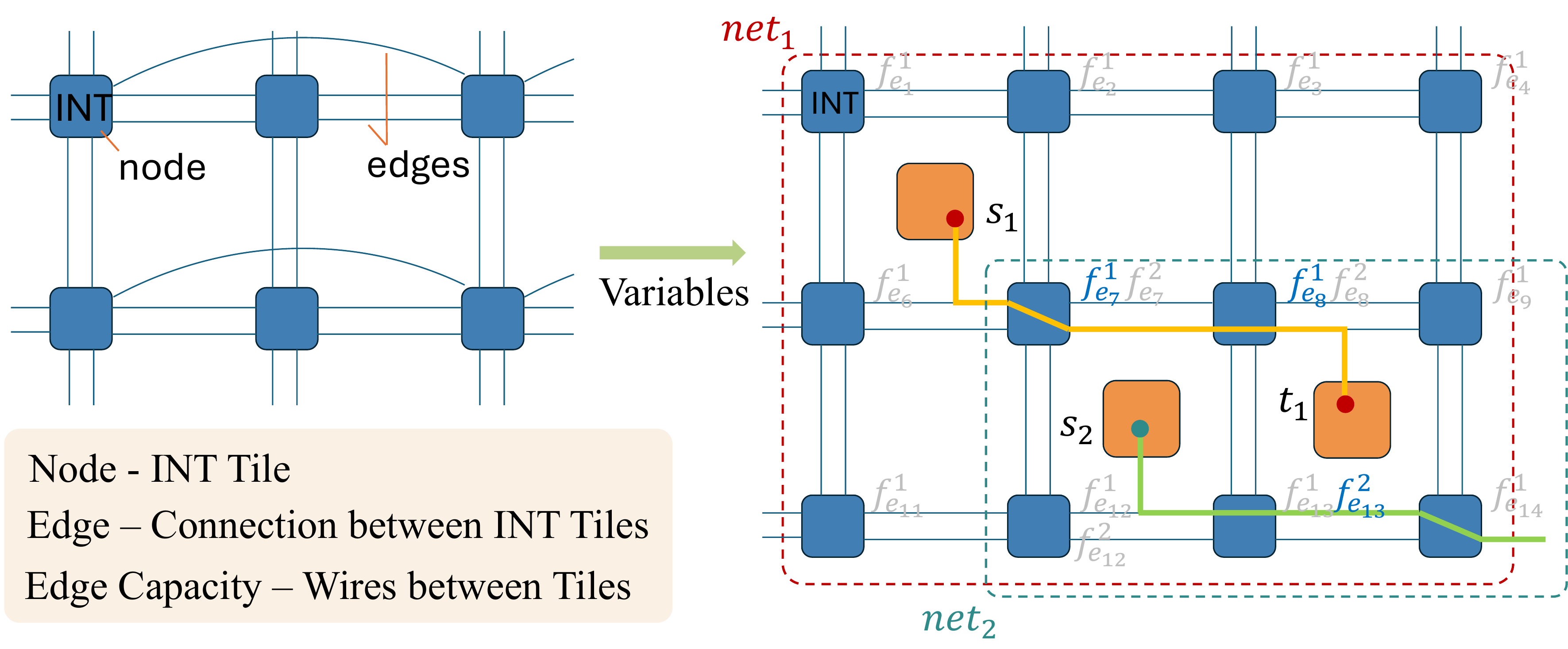}
    \caption{
    Simplified RRG and abstraction of routing variables
    }
    \label{fig:routing_variables}
\end{figure}

Figure~\ref{fig:routing_variables} illustrates the simplified RRG.
Each node encapsulates a cluster of internal switching resources, and each edge aggregates all feasible connections between tiles.
Edges form a 2D grid graph, capturing connectivity between adjacent tiles in the horizontal and vertical directions. 
In addition to the 2D grids, the RRG also includes longer-span edges that represent routing resources capable of traversing multiple tiles.

Based on the simplified RRG, a routing variable $f_i^k$ is defined as the probability that an edge $e_i$ of the RRG could be used by a net $n_k$.
The variable $f_i^k\in [0, 1]$ is a continuous variable that indicates the routing path of $n_k$ is highly likely to take the edge $e_i$ when it is close to $1$.

Routing typically exhibits significant spatial locality, which means the routing resources distant from pins of a net are largely irrelevant with the routing of the corresponding net.
The locality property allows the framework to prune redundant variables and constrain the search space.
After placement is applied, each net has fixed pin locations, and the pin locations form a bounding box.
The locality assumes that the routing path of a net rarely extends far beyond the bounding box.
Accordingly, we designate the routing region for each net as an expanded box of its bounding box~\cite{VPR8}.
A hyperparameter $\tau$ is defined to control the expansion ratio of the routing region to the bounding box.
As the INT tiles are arranged in a 2D array, the width and length are easily estimated by the number of INT tiles the bounding box spans.
For a bounding box of width $W_{n_k}$ and length $L_{n_k}$, the routing region will be expanded to $(1+\tau)W_{n_k}$ and $(1+\tau)L_{n_k}$ respectively.

The routing variables $\{f^k\}$ of net $n_k$ are restricted to the edges of the subgraph inside the designated routing region.
Figure~\ref{fig:routing_variables} demonstrates the abstraction of routing variables by an example of two nets.
The red box and blue are the routing regions for $net_1$ and $net_2$, respectively.
The variables $\{f^1_{e}\}$ of $net_1$ are defined over INT edges inside the red box, and the overlapping region between red and blue boxes encompasses variables for both $net_1$ and $net_2$.
The $\{f^1_{e_7}, f^1_{e_8}\}$ indicates the routing path for $net_1$, and $\{f^2_{e_13}, f^2_{e_14}\}$ indicates the routing path for $net_2$.
As variables $f$ are optimizable, defining differentiable objective functions over $f$ enables differentiable optimization for routing.

\subsection{Routing Objectives}
\label{sec:routing_obj}
The crucial challenge in differentiable routing implementation lies in modeling routing objectives and constraints with differentiable functions.
Following the definition of variables $f_i^k$ in Section~\ref{sec:preprocessing}, the framework models the routing problem in Equation~\ref{equ:routing_form}.
The abstract formulation for the differentiable optimization is as follows:

{\footnotesize
\begin{equation}
\label{equ:routing_form}
    \begin{aligned}
        \min_{f_i^k} \quad & WL(f) & (wirelength)\\
        s.t. \quad & 0 \leq \sum_{k} f_i^k \leq cap_{e_i} \quad for\;all\;edge\;e_i & (congestion) \\
        & n_k \text{ is connected}\quad for\;all\;net\;n_k & (connectivity) \\
    \end{aligned}
\end{equation}
}

\subsubsection{Wirelength} 

Wirelength is the most essential objective for routing.
Since the routing variables are defined directly on edge occupation, the wirelength of a net can be calculated as the sum of the lengths of the edges the net occupies.
We define the wirelength of net \(n_k\), denoted \(WL_{n_k}\), and the total wirelength \(WL_{\mathrm{tot}}\) as follows, where \(\alpha_{n_k}\) is the net importance/weight defined in Table~\ref{tab:notation}.

{
\footnotesize
\begin{equation}
    \label{equ:wirelength}
    \begin{aligned}
        WL_{tot} & = \sum \alpha_{n_k} \cdot (\sum_{e_i\in n_k}{f_i^k} \cdot w_{e_i})
    \end{aligned}
\end{equation}
}

\subsubsection{Connectivity}
A-star-based routing algorithms ensure the routing paths of each net are fully connected by searching for the next node from neighbor nodes of the current nodes.
Constrast to this path finding scheme, the differentiable routing solves the probability of each edge to be occupied by some net.
Therefore, additional constraints are required to ensure the edges occupied by the same net form a connected tree that spans all pins.

To model connectivity in a differentiable manner, we adopt a spectral perspective based on the graph Laplacian.
The graph Laplacian is formally defined as $L=D-A$ for a graph with adjacency matrix $A$ and degree matrix $D$.
Intuitively, the Laplacian matrix measures a form of smoothness across edges.
Spectral algorithm based on graph Laplacian can approximate the shortest path in a differentiable manner~\cite{MATH_JLAA2021_Steinerberger}, which proposes a criterion for the approximation by constructing a function $\phi: V\rightarrow \mathbb{R}_{+}$ for path between nodes $v_s$ and $v_u$:

{\footnotesize
\begin{equation}
    \label{equ:spectral_shortest_path}
    \begin{aligned}
        \phi = \arg\min_{\substack{g:V\rightarrow \mathbb{R}\\g(v_s)=0, g\not\equiv 0}} \frac{\sum_{<v_1, v_2>\in E}(g(v_1) - g(v_2))^2}{\sum_{v\in V}g(v)^2}
    \end{aligned}
\end{equation}
}

Equation~\ref{equ:spectral_shortest_path} indicates that $\phi$ is the assignment of function $g$ that minimizes the normalized quadratic form, where $g$ is a symbolic function.
The criterion of Equation~\ref{equ:spectral_shortest_path} provides guidance for the iterative construction of routing path based on the spectral heuristics.
To be more specific, the path could be constructed by gradually moving to the neighboring node with the smallest value of function $\phi$.

One explicit form of the function $\phi$ can be the smallest eigenvector of the graph Laplacian matrix with the columns and rows corresponding to node $v_s$ removed.
Inspired by the spectral approach to the shortest path, we adapt the differentiable objective for net connectivity from the algebraic connectivity~\cite{MATH_CMJ1973_Fiedler, MATH_LMA2005_Wu}.
For each net $n_k$, the Laplacian matrix is defined separately based on the subgraph introduced in Section~\ref{sec:preprocessing}.
The raw Laplacian matrix is defined as $L^{n_k} = D^{n_k} - A^{n_k}$, where $D^{n_k}$ is the degree matrix and $A^{n_k}$ is the adjacency matrix.
Different from the $0,1$-connection induced adjacency matrix, $A^{n_k}$ is weighted by $f^k_i$.
The derived Laplacian matrix $L_{\backslash P_{n_k}}^{n_k}$ is defined as removing the corresponding columns and rows of the pin nodes $P_{n_k}$ from $L^{n_k}$.
The non-connectivity penalty $CV_{n_k}$ on net $n_k
$ represents the differentiable constraint.
The non-connectivity penalty $CV_{n_k}$ as algebraic connectivity is defined as follows:

{\footnotesize
\begin{equation}
    \label{equ:non_connectivity}
    \begin{aligned}
        CV_{n_k} &= -\lambda_2(L^{n_k}_{\backslash P_{n_k}}) \\
    \end{aligned}
\end{equation}
}

Computing the gradient of the function $CV_{n_k}$ in Equation~\ref{equ:non_connectivity} is computationally expensive.
We therefore introduce the approximation method to the gradient computation in Section~\ref{sec:grad_approx}.

\subsubsection{Congestion}
Conventional routing algorithms resolve congestion between nets by rip-up and re-route strategy.
In constrast, the differentiable formulation addresses congestion proactively during the optimization process, jointly considering congestion alongside other routing objectives.
There is a duality between the routing problem and the multi-commodity flow problem~\cite{iccd2007hu}.
Inspired by the flow problem, we formulate the problem of managing congestion as constraints on the usage of each edge and conservation on each node.

\begin{figure}[tb]
    \centering
    \includegraphics[width=0.86\columnwidth]{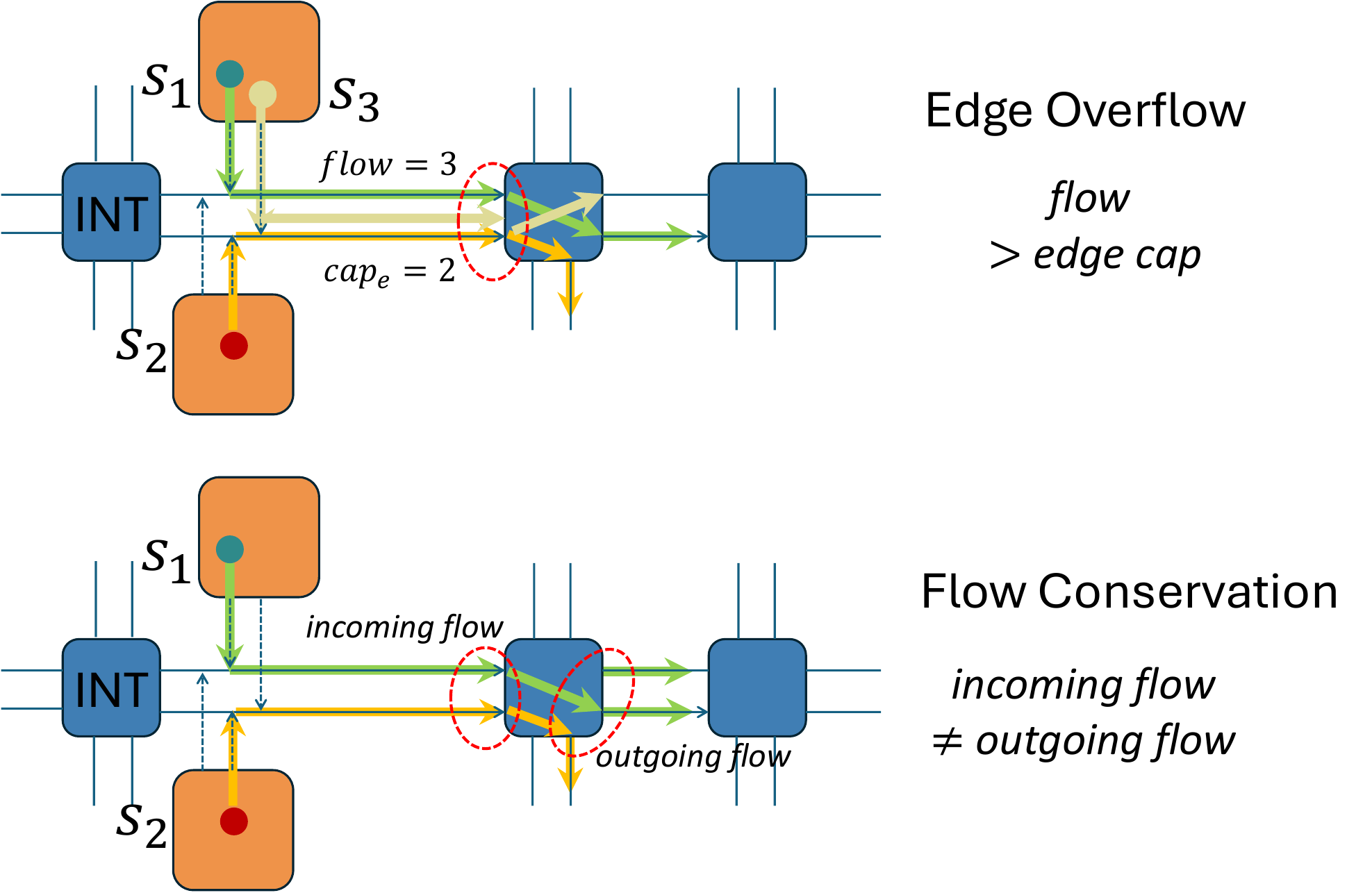}
    \caption{
    Example of edge overflow and flow conservation}
    \label{fig:edge_node_overflow}
\end{figure}

Edge overflow and flow conservation are the two main criteria for reducing congestion.
The overflow on an edge compared to the edge capacity $cap_{e}$ can be calculated by the accumulated usage of different nets over this edge.
The edge capacity $cap_{e}$ is intrinsic, determined by the PIP resources between tiles.
The flow conservation requires the incoming flow and outgoing flow of the same node to be equal.
Except for the nodes related to pins, all nodes are supposed to satisfy the flow conservation requirements to avoid redundancy of edge usage. 

Figure~\ref{fig:edge_node_overflow} demonstrates an example of edge overflow and flow conservation.
There are three nets in the edge overflow case. 
The three nets take the same edge of capacity $cap_e=2$, which means an overflow is triggered.
In the flow conservation case, there are two nets using the same INT tile.
Overflow occurs when the incoming flows are $2$, but the outgoing flows are $3$.
To generate feasible routing solutions, capacity violations on edges and inequality on nodes are penalized during optimization by defining two objective functions $OVF_{e}$ and $OVF_{v}$.
The $OVF_{e_i}$ function on one single edge $e_i$ and the $OVF_{v_j}$ function on one single node $v_j$ are defined respectively as follows:

{\footnotesize
\begin{equation}
    \label{equ:edge_exclusive}
    \begin{aligned}
        OVF_{e_i} & = ReLU(\sum_{n_k}f_i^k - cap_{e_i}) \\
        OVF_{v_j} & = (\sum_{e_i\in E^{-}(v_j)} \sum_{n_k} f_i^k - \sum_{e_i\in E^{+}(v_j)} \sum_{n_k} f^k_i)^2
    \end{aligned}
\end{equation}
}

\subsection{Optimizer}
\label{sec:optimizer}

Constraints of differentiable optimization can be relaxed to the objective function in practice, to have a better trade-off between solution quality and computational efficiency.
The original formulation considers constraints of congestion and connectivity in Equation~\ref{equ:routing_form}.
Net connectivity is enforced via the objective function defined in Equation~\ref{equ:non_connectivity}.
The edge usage constraint can be represented as $OVF_{e_i}=0$ for each edge $e_i$ and $OVF_{v_j} = 0$ for each node $v_j$ from Equation~\ref{equ:edge_exclusive}.
Accordingly, the routing formulation in Equation~\ref{equ:routing_form} can be rewritten using the augmented Lagrangian method as follows:

{\footnotesize
\begin{equation}
    \label{equ:relaxed_opt}
    \begin{aligned}
        \min_{f_i^k} \quad & WL_{tot} + \mu CV + \sum_{s\in \mathcal{S}}\eta_s(OVF_s + \frac{c_s}{2}OVF_s^2) \\
        s.t. \quad & CV = \sum_{n_k} CV_{n_k},\quad \mathcal{S} = \{E, V\} \\
    \end{aligned}
\end{equation}
}

where $\mu$ and $\eta_s$ are the multipliers to regulate the convergence behavior, and $c_s$ denotes the weight of the quadratic penalty term $OVF_s^2$.
The term $s$ could be the edge set $E$ or the node set $V$.
Throughout the optimization process, multipliers $\mu$ and $\eta$ are adaptively updated using the normalized subgradient method from elfPlace~\cite{FPGA_ICCAD2019_Li}.

We employ the ADAM optimizer to solve the optimization form of Equation~\ref{equ:relaxed_opt}.

\begin{algorithm}[tb]
    \caption{Approximation to $CV_{n_k}$ Gradient}
    \label{alg:grad_cv}
    \begin{algorithmic}[1]
        \Require The net $n_k$ and Laplacian $L^{n_k}_{\backslash P_{n_k}}$
        \Ensure The gradient of $CV_{n_k}$ to $f_i^k$
        \Function{ApproxGrad}{$m$}
        \State init $\beta_0 \gets 0$, $\mathbf{q_0}\gets \mathbf{0}$
        \For{iteration $t \gets 1\cdots m$}
        \State $\mathbf{\gamma} \gets L^{n_k}_{\backslash P_{n_k}} \mathbf{q_t}$, $\alpha_t \gets \mathbf{q_t} \cdot \mathbf{\gamma}$
        \State $\mathbf{\gamma} \gets \mathbf{\gamma} -  \alpha_t \mathbf{q_t} - \beta_{t-1} \mathbf{q_{t-1}}$
        \State $\beta_t\gets \vert \gamma \vert $, $\mathbf{q_{t+1}}\gets \gamma / \beta_t$
        \EndFor
        \State form tridiagonal matrix $T$ from $\{\alpha_t, \beta_t\}$
        \State form matrix $Q$ from $\mathbf{q_t}$
        \State compute eigenpair $\theta_2, \mathbf{y_2}$ of $T$
        \State $\mathbf{u} \gets Q \mathbf{y_2}$, $\mathbf{v_2} \gets \mathbf{u} / \Vert \mathbf{u}\Vert$
        \For{$e_i \in E$}
        \State $e_i$ connect nodes $v_i$ and $v_j$
        \State $Grad(f^k_i)\gets (\mathbf{v_2}[v_i] - \mathbf{v_2}[v_j])^2$
        \EndFor
        \State \Return $Grad(f^k_i)$ 
        \EndFunction
    \end{algorithmic}
\end{algorithm}

\subsubsection{Gradient computation}
\label{sec:grad_approx}
The gradient computation for the wirelength term $WL$ and the overflow term $OVF$ is straightforward.
In contrast, directly computing the gradient of the non-connectivity penalty term $CV$ can incur high computational costs.
The gradient of $CV_{n_k}$ with respect to the variable $f_i^k$ basically depends on the eigenvector $\mathbf{v_2}$ of the derived Laplacian matrix $L_{\backslash P_{n_k}}^{n_k}$.
Specifically, the gradient $f_i^k$ is determined by the eigenvector elements at the corresponding position of nodes incident to edge $e_i$.
The gradient can be explicitly computed as follows:

{\footnotesize
\begin{equation}
    \label{equ:grad_cv}
    \begin{aligned}
        \frac{\partial CV_{n_k}}{\partial f^k_i} = \mathbf{v_2}^T \frac{\partial L^{n_k}_{\backslash P_{n_k}}}{\partial f^k_i}\mathbf{v_2}
    \end{aligned}
\end{equation}
}

Computation of the explicit gradient value by Equation~\ref{equ:grad_cv} could be prohibitively expensive for an optimization iteration.
Lanczos algorithm provides an efficient method to approximate eigenvalues and eigenvectors~\cite{MATH_JRNBS1950_Lanczos}.
We employ the iterative scheme of the Lanczos algorithm to approximate the gradients defined in Equation~\ref{equ:grad_cv}.
Algorithm~\ref{alg:grad_cv} depicts the detailed steps.
The key steps of Algorithm~\ref{alg:grad_cv} (line 4-8) are compatible with tensor operations.
We implement the GPU acceleration of Algorithm~\ref{alg:grad_cv} to further improve computational efficiency.

\subsection{Postprocessing}
\label{sec:legalization}

The routing stage generates initial routing solutions with simplified RRG and relaxed constraints of connectivity and congestion. 
Therefore, we introduce an additional postprocessing stage to finalize the routing solutions.
There are two challenges for generating the final routing solutions:
1) The routing paths of a net could not be fully connected;
2) routing solutions include redundant edges.

We denote the initial solutions for the net $n_k$ from routing stage as a set of edges $E^k$.
The fully connected net routing of net $n_k$ is a superset of $E_k$ that serves as the minimum spanning tree of $n_k$.
If an edge can be removed without disconnecting pins, the edge is redundant and can be pruned.

\begin{algorithm}[tb]
    \caption{Processing Initial Routing Solutions}
    \label{alg:legalization}
    \begin{algorithmic}[1]
        \Require Initial routing solutions $E^k$ for net $n_k$ 
        \Ensure Routing solution $\bar{E^k}$ for net $n_k$
        \Function{Legalize}{$\{E^k\}$}
        \For{net $n_k$}
        \State $\{C_j\}\gets$\Call{ConnectedComponents}{$E^k$}
        \State $dist\gets$\Call{ShortestPath}{$\{C\}$}
        \State $G^C\gets \langle V^C, E^C, dist\rangle$ 
        \State $\tilde{E^k}\gets E^k \cup$ \Call{MST}{$G^C$}
        \State $\Pi \gets$ \Call{PostOrderDFS}{$\tilde{E^k}, p_r$}
        \For{$e_i\in \Pi$}
        \If{$\nexists$ pin $p\in$\Call{SubTree}{$e_i$}}
        \State $\tilde{E^k}\gets \tilde{E^k}-\{e_i\}$
        \EndIf
        \EndFor
        \EndFor
        \State \Return $\{\tilde{E^k}\}$
        \EndFunction
    \end{algorithmic}
\end{algorithm}

Algorithm~\ref{alg:legalization} demonstrates main steps for legalizing the initial routing solutions for a net.
For a net $n_k$, the algorithm first finds all connected components $\{C_j\}$ from the initial routing solution $E^k$ (line 3).
The algorithm computes the minimum distances between all components $C$ and stores the actual shortest paths (line 4).
A complete graph $G^C$ over components considers each component as a graph node and the edge weight between 2 nodes to be the minimum distance between the corresponding components (line 5).
The algorithm run minimum spanning tree algorithm on the component graph $G^C$ and extend the routing solutions with the paths added for spanning tree (line 6).
To remove redundant edges, the algorithm executes depth-first search on the spanning tree $\tilde{E^k}$ and records the information whether there is a pin in the subtree of the edge (line 7).
Without a pin in subtree, the redundant edges are removed from the spanning tree (line 8-10).
The spanning trees $\{\tilde{E^k}\}$ for all nets are constructed from well-optimized initial solutions and have only a few congestions or violations.

The solutions $\{\tilde{E^k}\}$ will be the guidance for the final routing solutions in the manner of global routing and detailed routing.
We rely on the partitioning technique~\cite{FPGA_GLSVLSI2024_Zang} to call single-net routing and adapt the RapidWright routing kernel~\cite{FPGA_TRETS2021_Zhou} for single-net routing.
For each net $n_k$, $\{\tilde{E^k}\}$ indicates a set of INT tiles and INT-tile-connections, so the algorithm can implicitly map the initial solution $\{\tilde{E^k}\}$ to routing regions by checking a PIP attribute based on tile location when the routing kernel explores the PIP as a candidate.
The routing kernel increases the routing cost outside regions with a penalty function linear to the least distance from the PIP to tiles of the routing region.
Potential conflicts are resolved by the rip-up and reroute technique.

\section{Experiments} 
\label{sec:exp}

Our routing framework \textit{DiffRouter} is implemented in Python with PyTorch, Triton, and RapidWright Jython integration.
The experiments are conducted on a Linux server with AMD EPYC 9554P CPU @3.1GHz 
with 128 cores and 755GB RAM
, and Nvidia RTX A6000 GPU.

\subsection{Experimental Setup}

\subsubsection{Benchmarks} 

We study the efficiency of our routing framework on 2024 Runtime-First FPGA Interchange Routing Contest (FPGA 2024)~\footnote{\url{https://xilinx.github.io/fpga24_routing_contest/}}.
The benchmarks consist of 28 circuits from 12 different source suites.
The circuits are synthesized from RTL designs, placed, and routed clock nets with Vivado~\cite{Vivado2023_2} by the contest organizer.
In these circuits, the range of the numbers of LUTs, FFs, DSPs, and BRAMs are 31k-344k, 2k-399k, 0-2209, and 0-720, respectively.
The benchmarks target the AMD-Xilinx UltraScale+ xcvu3p device.

The contest officially evaluates routing results by two metrics: program runtime and critical path wirelength (CPWL). 
CPWL is defined as the total wirelength of the longest path in the routing and is computed as the sum of the lengths of all PIPs used to connect source and sinks.
In addition, we report total wirelength (tot WL) as an auxiliary routing-quality metric. Tot WL is \emph{not} an official contest metric, and the contest does not provide an official tot WL analyzer. Tot WL is defined as the total wirelength over all nets and is computed by our calculator derived from the official CPWL analyzer, which applies the same wirelength convention to all routers. 
As for the runtime, we adopt wall time instead of CPU time.
The top-three participants of the contest have made their papers and results publicly available~\cite{FPGA_FPL2024_Martin, FPGA_GLSVLSI2024_Zang, AceRoute}.

\subsubsection{Baselines} We compare our routing framework with 4 baselines described as follows:
\begin{enumerate}
    \item Vivado: we utilize Vivado v2023.2~\cite{Vivado2023_2} router with non-timing-driven mode as a baseline.
    \item RWRoute~\footnote{\url{https://github.com/Xilinx/RapidWright}}: RWRoute is an open-source FPGA router integrated in RapidWright framework~\cite{FPGA_TRETS2021_Zhou}, which is the baseline router for FPGA 2024 contest.
    \item CUFR~\footnote{\url{https://github.com/xszang/parallel-routing}}: CUFR is the second place and the highest-ranked open-source router in FPGA 2024 contest~\cite{FPGA_GLSVLSI2024_Zang}. CUFR builds a partitioning tree to utilize CPU parallelism. We set the number of threads to 32, consistent with the best reported results. 
    \item Potter~\footnote{\url{https://github.com/diriLin/Potter}}: Potter is the state-of-the-art open-source router targeted at runtime on FPGA 2024 benchmarks~\cite{FPGA_ICCAD2025_Zang}. The number of threads is also set to 32, aligned with CUFR and the best reported results.
\end{enumerate}

\subsection{Experimental Results}

Table~\ref{tab:exp_res} reports the overall performance of \textit{DiffRouter} on the FPGA 2024 benchmarks, compared against Vivado~\cite{Vivado2023_2}, RWRoute~\cite{FPGA_TRETS2021_Zhou}, CUFR~\cite{FPGA_GLSVLSI2024_Zang}, and Potter~\cite{FPGA_ICCAD2025_Zang}. 
Vivado fails to route three designs: \texttt{rosetta\_fd}, \texttt{corundum\_25g}, and \texttt{corundum\_100g}. 
Runtime is measured as end-to-end execution time (seconds).
CPWL denotes the critical-path wirelength reported by the contest analyzer, and tot WL denotes the total wirelength computed by our calculator.
The last row shows the geometric mean of per-benchmark ratios, normalized to \textit{DiffRouter} (set to 1.0).

\textit{DiffRouter} improves both efficiency in most benchmarks and routing quality in part of benchmarks. 
Compared to RWRoute, \textit{DiffRouter} achieves a \(5\times\) speedup while delivering \(5.3\%\) lower CPWL. 
It is also approximately \(2\times\) faster than Vivado on the benchmarks that Vivado successfully routes, while achieving \(42.6\%\) lower CPWL. 
Against the strongest academic baseline, Potter, \textit{DiffRouter} reduces runtime by \(18.2\%\) on average, with up to a \(42.7\%\) reduction on \texttt{boom\_soc}. 
Overall, \textit{DiffRouter} attains the best CPWL among all baselines and the lowest tot WL among the open-source routers, while Vivado achieves lower tot WL on the designs it routes successfully. These results demonstrate that our differentiable routing formulation can simultaneously improve parallel efficiency beyond state-of-the-art PathFinder-based routers, even when those baselines already exploit parallelism.
We next present ablation studies to isolate the key optimization settings and design choices that drive these gains.

\begin{table*}[tb]
\centering
\caption{Comparative results on FPGA 2024 Contest benchmarks.}
\begin{threeparttable}
\resizebox{0.99\textwidth}{!}{
\begin{tabular}{lrrrrr rrrrr rrrrr}
    \toprule
    & \multicolumn{5}{c}{\textbf{Runtime (s)}} & \multicolumn{5}{c}{\textbf{CPWL}} & \multicolumn{5}{c}{\textbf{tot WL ($\times10^6$)}} \\
    \cmidrule(lr){2-6} \cmidrule(lr){7-11} \cmidrule(lr){12-16}
    \textbf{Benchmark}
    & \cite{Vivado2023_2}
    & \cite{FPGA_TRETS2021_Zhou}
    & \cite{FPGA_GLSVLSI2024_Zang}
    & \cite{FPGA_ICCAD2025_Zang}
    & \textit{DiffRouter}
    & \cite{Vivado2023_2}
    & \cite{FPGA_TRETS2021_Zhou}
    & \cite{FPGA_GLSVLSI2024_Zang}
    & \cite{FPGA_ICCAD2025_Zang}
    & \textit{DiffRouter} 
    & \cite{Vivado2023_2}
    & \cite{FPGA_TRETS2021_Zhou}
    & \cite{FPGA_GLSVLSI2024_Zang}
    & \cite{FPGA_ICCAD2025_Zang}
    & \textit{DiffRouter}
    \\
    \midrule
    \texttt{logicnets\_jscl} & 26 & 51 & 36 & 25 & 31 & 290 & 217 & 234 & 221 & 223 & 0.412 & 0.499 & 0.510 & 0.580 & 0.501 \\
    \texttt{boom\_med\_pb} & 63 & 221 & 181 & 95 & 65 & 820 & 955 & 806 & 817 & 823 & 1.384 & 1.884 & 1.782 & 1.968 & 1.719 \\
    \texttt{vtr\_mcml} & 270 & 232 & 111 & 58 & 60 & 614 & 570 & 584 & 607 & 578 & 1.419 & 1.668 & 1.659 & 1.832 & 1.651 \\
    \texttt{rosetta\_fd} & -- & 147 & 139 & 61 & 45 & -- & 839 & 804 & 812 & 810 & - & 1.858 & 1.834 & 2.029 & 1.850 \\
    \texttt{corundum\_25g} & -- & 256 & 135 & 72 & 61 & -- & 442 & 500 & 401 & 421 & - & 3.898 & 3.913 & 4.242 & 3.815 \\
    \texttt{finn\_radioml} & 65 & 118 & 63 & 35 & 37 & 341 & 272 & 251 & 258 & 223 & 1.350 & 1.536 & 1.545 & 1.772 & 1.545 \\
    \texttt{vtr\_lu64peeng} & 120 & 215 & 130 & 67 & 60 & 1607 & 1342 & 1333 & 1308 & 1345 & 2.675 & 3.049 & 3.081 & 3.458 & 3.061 \\
    \texttt{corescore\_500} & 82 & 161 & 75 & 40 & 41 & 771 & 660 & 668 & 655 & 660 & 1.957 & 2.220 & 2.218 & 2.606 & 2.248 \\
    \texttt{corescore\_500\_pb} & 105 & 262 & 141 & 80 & 69 & 940 & 709 & 739 & 660 & 671 & 2.621 & 3.318 & 3.379 & 3.643 & 3.224 \\
    \texttt{mlcad\_d181\_lefttwo3rds} & 197 & 1761 & 439 & 188 & 180 & 989 & 868 & 771 & 727 & 714 & 8.757 & 10.346 & 9.888 & 10.634 & 9.791 \\
    \texttt{koios\_dla\_like\_large} & 210 & 505 & 206 & 92 & 75 & 831 & 545 & 520 & 552 & 536 & 7.077 & 7.467 & 7.507 & 8.188 & 7.455 \\
    \texttt{boom\_soc} & 443 & 2068 & 792 & 351 & 201 & 2204 & 1717 & 1673 & 1708 & 1711 & 7.562 & 9.168 & 9.151 & 10.302 & 9.080 \\
    \texttt{ispd16\_example2} & 195 & 645 & 324 & 172 & 109 & 1437 & 1014 & 939 & 1003 & 951 & 8.621 & 10.062 & 10.159 & 11.261 & 10.048 \\
    \texttt{rapidwright\_picoblazearray} & 93 & 173 & 67 & 29 & 33 & 144 & 120 & 111 & 110 & 108 & 1.871 & 1.951 & 1.934 & 2.207 & 1.962 \\
    \texttt{corundum\_100g} & -- & 224 & 100 & 59 & 50 & -- & 550 & 549 & 562 & 566 & - & 4.253 & 4.272 & 4.640 & 4.205 \\
    \texttt{koios\_clstm\_like\_large} & 113 & 210 & 99 & 48 & 48 & 411 & 254 & 228 & 238 & 233 & 3.385 & 3.464 & 3.474 & 3.856 & 3.467 \\
    \texttt{titan23\_orig\_gsm\_x6} & 207 & 443 & 177 & 92 & 61 & 488 & 275 & 289 & 271 & 280 & 5.374 & 6.004 & 5.924 & 6.517 & 5.841 \\
    \texttt{corescore\_900} & 157 & 312 & 165 & 66 & 63 & 1590 & 1176 & 1216 & 1215 & 1205 & 3.612 & 4.140 & 4.149 & 4.907 & 4.199 \\
    \texttt{koios\_dla\_like\_large\_v2} & 224 & 481 & 265 & 88 & 70 & 553 & 341 & 284 & 311 & 307 & 8.683 & 9.086 & 9.057 & 9.815 & 9.053 \\
    \texttt{finn\_mobilenetv1} & 271 & 475 & 229 & 96 & 63 & 542 & 360 & 320 & 384 & 359 & 4.380 & 5.003 & 5.080 & 5.933 & 5.054 \\
    \texttt{ispd16\_fpga03} & 239 & 609 & 367 & 188 & 149 & 1629 & 859 & 848 & 873 & 823 & 7.964 & 8.895 & 9.001 & 9.909 & 8.931 \\
    \texttt{mlcad\_d181} & 352 & 4433 & 671 & 344 & 292 & 1011 & 1064 & 828 & 779 & 774 & 13.759 & 16.577 & 15.761 & 16.932 & 15.582 \\
    \texttt{boom\_soc\_v2} & 449 & 4142 & 1198 & 416 & 282 & 1842 & 1191 & 1020 & 982 & 1027 & 9.984 & 12.799 & 12.431 & 13.422 & 12.071 \\
    \texttt{corescore\_1200} & 214 & 468 & 180 & 84 & 86 & 1442 & 1047 & 1105 & 1106 & 1127 & 5.063 & 5.762 & 5.773 & 6.781 & 5.845 \\
    \texttt{ispd16\_example2\_v2} & 211 & 566 & 289 & 124 & 94 & 1242 & 728 & 683 & 669 & 689 & 7.517 & 8.666 & 8.726 & 9.719 & 8.646 \\
    \texttt{corescore\_1500} & 283 & 731 & 288 & 147 & 129 & 1597 & 1301 & 1331 & 1302 & 1315 & 6.507 & 7.739 & 7.790 & 9.411 & 7.892 \\
    \texttt{titan23\_orig\_dart\_x4} & 360 & 848 & 253 & 118 & 114 & 1049 & 634 & 567 & 639 & 629 & 6.849 & 8.115 & 8.181 & 9.185 & 8.004 \\
    \texttt{corescore\_1700} & 493 & 1591 & 669 & 231 & 169 & 2065 & 1595 & 1544 & 1570 & 1576 & 8.921 & 11.889 & 12.067 & 12.897 & 11.452 \\
    \midrule
    Geometric mean (ratio) & 2.112 & 5.475 & 2.488 & 1.184 & \textbf{1.000} & 1.426 & 1.053 & 1.007 & 1.008 & \textbf{1.000} & 0.870 & 1.012 & 1.010 & 1.127 & \textbf{1.000} \\
    \bottomrule
\end{tabular}
}
\begin{tablenotes}
\footnotesize
\item 1. Vivado fails to route three designs: \texttt{rosetta\_fd}, \texttt{corundum\_25g}, and \texttt{corundum\_100g};
\item 2. Baselines: Vivado~\cite{Vivado2023_2}, RWRoute~\cite{FPGA_TRETS2021_Zhou}, CUFR~\cite{FPGA_GLSVLSI2024_Zang}, Potter~\cite{FPGA_ICCAD2025_Zang};
\item 3. Tot WL is an auxiliary metric, not an official contest metric. The tot WL columns are corrected in this version (arXiv v2); see the Correction Notice.
\end{tablenotes}
\end{threeparttable}
\label{tab:exp_res}
\end{table*}

\subsection{Runtime Breakdown}

For large benchmarks, we analyze the runtime breakdown of \textit{DiffRouter} to better understand the routing performance.

\begin{figure}[tb]
    \centering
    \includegraphics[width=0.46\textwidth]{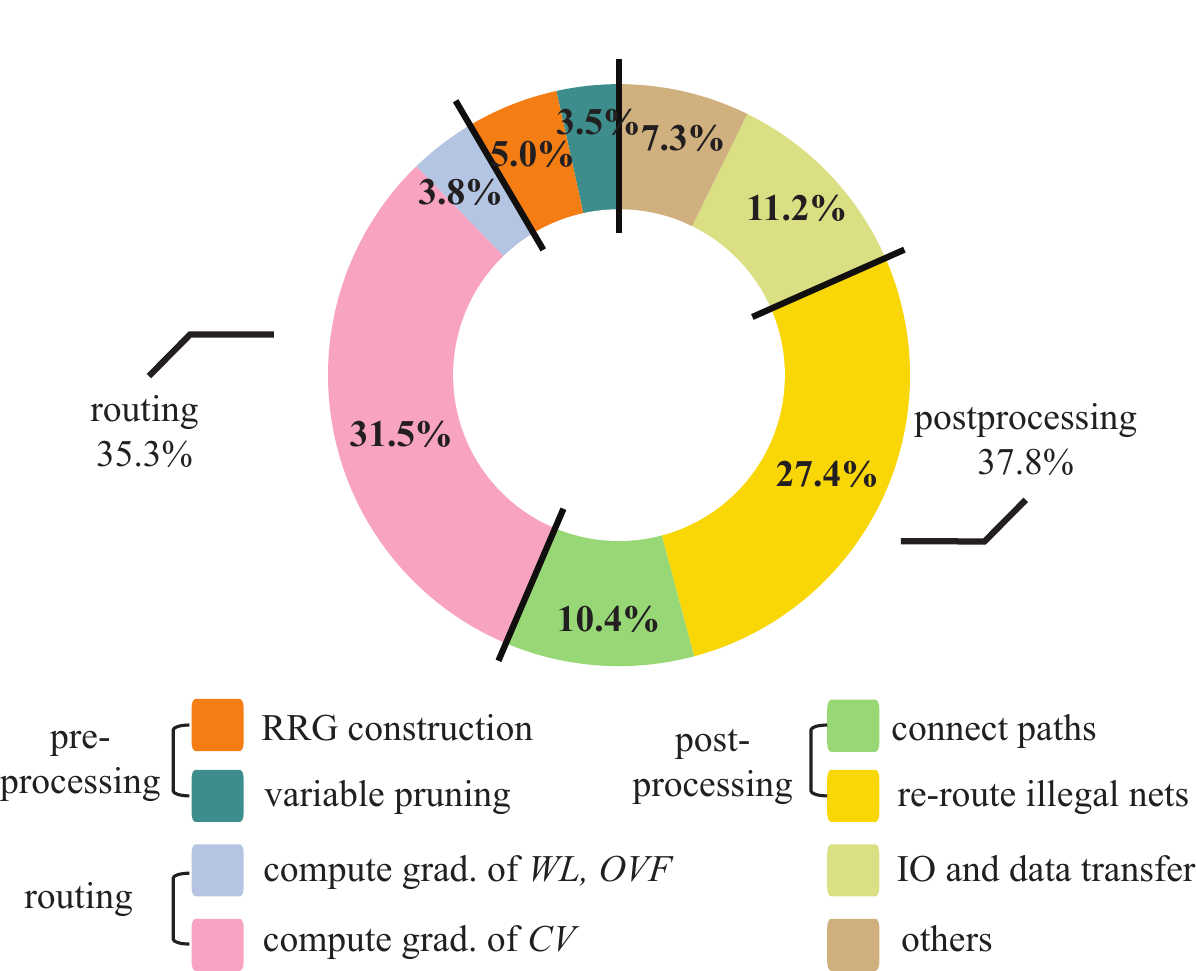}
    \caption{The runtime breakdown of \textit{DiffRouter} on benchmark \texttt{boom\_soc\_v2}}
    \label{fig:rtbd}
\end{figure}

Fig.~\ref{fig:rtbd} shows the runtime breakdown of \textit{DiffRouter} on benchmark \texttt{boom\_soc\_v2}, which is a relatively large benchmark.
\textit{DiffRouter} has three main stages of preprocessing, routing, and postprocessing.
For benchmark \texttt{boom\_soc\_v2}, the routing stage accounts for 35.3\% of the total runtime.
Within the routing stage, the computation of the gradient of the connectivity function is the most computationally intensive task.
Notably, the routing stage is GPU-accelerated with $50\times$ speed-up compared to CPU computation due to the highly parallelizable tensor operations.
The postprocessing stage constitutes the largest portion of the execution runtime, and re-routing all illegal nets occupies the most runtime.

We observe that benchmarks with similar sizes tend to exhibit similar runtime distributions. 
We categorize the benchmarks to small and large based on their size, using 150k LUTs as a threshold. 
For small benchmarks, pre-processing, routing, post-processing, and others account for approximately 10.4\%, 36.7\%, 29.8\%, and 23.1\% of the total runtime on average, while for large benchmarks, pre-processing, routing, post-processing, and others account for approximately 7.8\%, 34.9\%, 42.3\%, and 15.0\% of the total runtime on average.

\subsection{Ablation Study}

We design two ablation experiments to evaluate how the efficiency of our framework benefits from the optimization setup, including the number of iterations and the portion of nets routed at the differentiable routing stage.
We take suite \texttt{boom} from FPGA 2024 Contest benchmarks for example and conduct ablation experiments on the benchmarks: \texttt{boom\_med\_pb}, \texttt{boom\_soc}, and \texttt{boom\_soc\_v2}.

Before presenting the ablation results, we first estimate how close the routing stage is to feasibility by examining how many edges are overused.
An edge is considered overused if the total number of nets using the edge exceeds its capacity.
Table~\ref{tab:ablation_overused_edges} shows the proportions of overused edges, and the overused edges are further categorized based on the demand over capacity to distinguish between slight and severe over-usage.
Specifically, $>100\%$ indicates edge usage exceeds capacity, and $>200\%$ indicates $200\%$ exceeding $100\%$ more than capacity.
The number of overused edges is relatively small, and the legality issue will be resolved in the post-processing stage.

\begin{table}[tb]
\centering
\caption{
Overused edge proportion after routing stage.
}
\label{tab:ablation_overused_edges}
\resizebox{0.84\linewidth}{!}{
\begin{tabular}{lrrrr}
\toprule
\textbf{Benchmark} & $>100\%$ & $>120\%$ & $>150\%$ & $>200\%$ \\
\midrule
\texttt{boom\_med\_pb} & 2.7\textperthousand & 1.2\textperthousand & 0\textperthousand & 0\textperthousand \\
\texttt{boom\_soc} & 10.1\textperthousand & 6.5\textperthousand & 1.4\textperthousand & 0\textperthousand \\
\texttt{boom\_soc\_v2} & 11.1\textperthousand & 6.6\textperthousand& 2.1\textperthousand& 0\textperthousand\\
\bottomrule
\end{tabular}
}
\end{table}

\subsubsection{Optimization iterations}
The differentiable routing stage terminates the iterative process upon satisfying stable convergence.
We analyze how different numbers of optimization iterations affect the overall runtime and CPWL results.

\begin{figure}[tb]
    \centering
    \includegraphics[width=0.46\textwidth]{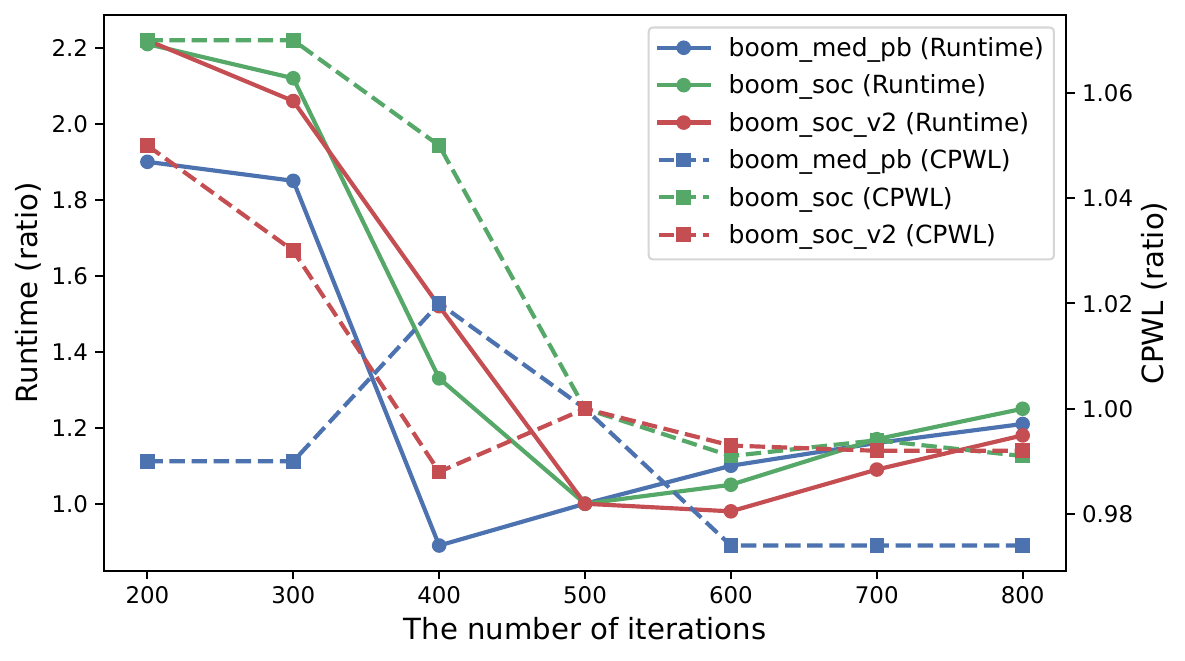}
    \caption{Runtime and CPWL sensitivity of iterations}
    \label{fig:ablation1}
\end{figure}

Fig.~\ref{fig:ablation1} illustrates the sensitivity of runtime and CPWL to the number of optimization iterations on suite \texttt{boom} benchmarks.
As the number of iterations increases, the runtime of the differentiable routing stage increases, while the runtime of post-processing could possibly decrease.
The overall runtime of the framework reaches a minimum mostly at iterations between 400 and 600.
As for CPWL, increasing iterations slightly improves solution quality.
The marginal CPWL reduction becomes limited beyond 600 iterations for benchmarks \texttt{boom\_soc} and \texttt{boom\_soc\_v2}.

\begin{figure}[tb]
    \centering
    \includegraphics[width=0.45\textwidth]{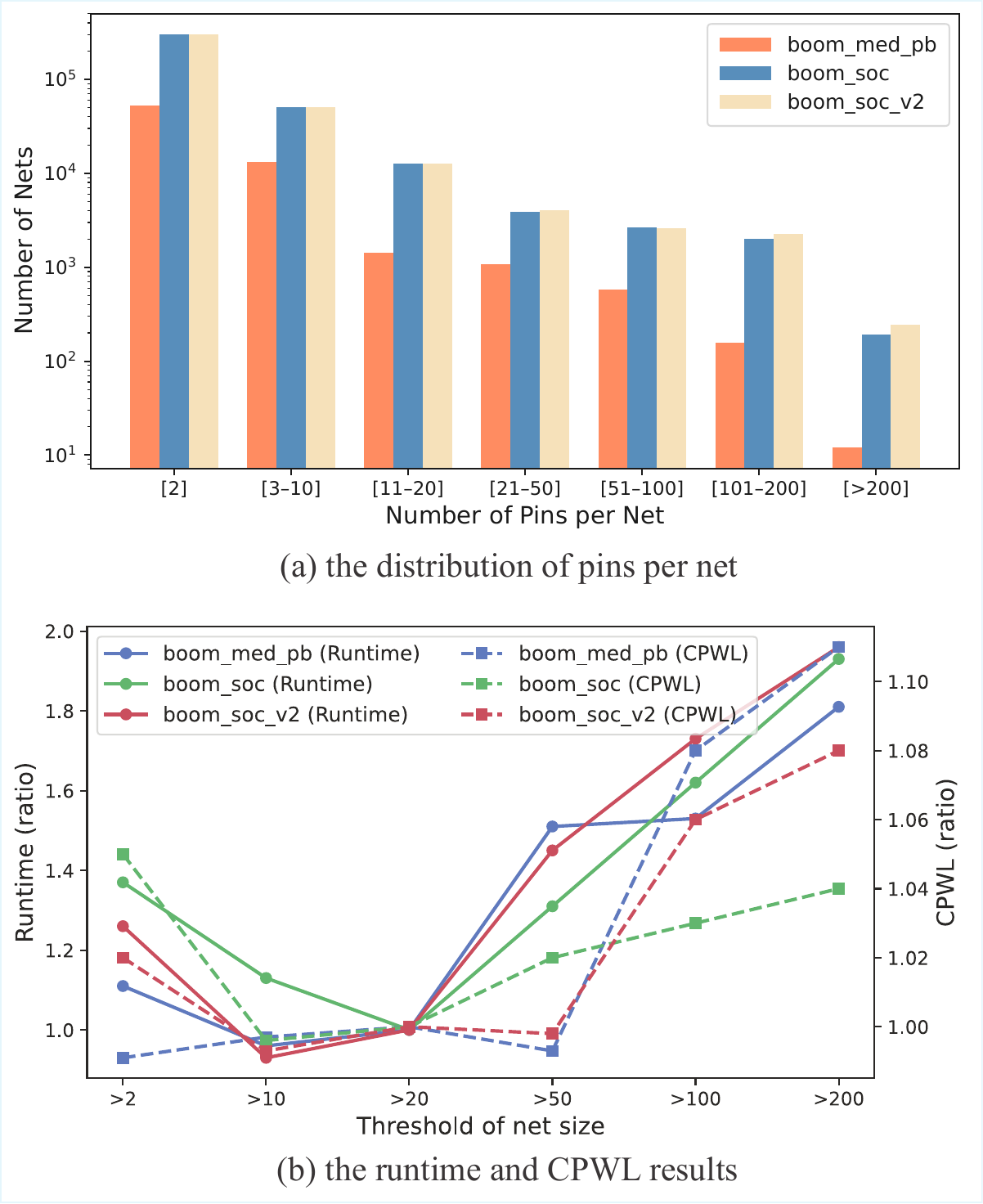}
    \caption{Runtime and CPWL sensitivity of net size threshold}
    \label{fig:ablation2}
\end{figure}

\subsubsection{Threshold of net size}
Our framework considers routing a part of nets at the differentiable routing stage.
The differentiable routing stage can focus on processing high-fanout nets.
A simple heuristic is to define a threshold for the number of pins (net size).
The differentiable routing stage specifically targets those nets that surpass the predefined threshold.

Fig.~\ref{fig:ablation2} (a) illustrates the distribution of net sizes of the three benchmarks from suite \texttt{boom}.
Nets are categorized by the number of pins, and the y-axis represents the number of nets in each category on a logarithmic scale.
The majority of nets have small fanout (mostly 2-20 pins), and the amount of high-fanout nets are usually less than 10\%.
Nevertheless, high-fanout nets are more prone to inducing routing congestion and tend to consume a disproportionate amount of routing resources in sequential routing methods.

Fig.~\ref{fig:ablation2} (b) demonstrates the runtime and CPWL sensitivity of the net size threshold.
The x-axis represents the threshold of net size.
For instance, the ``$>20$'' indicates that only nets with a net size exceeding 20 are processed during the differentiable routing stage.
The runtime and CPWL results are plotted on the left and right y-axes, respectively.
The values of y-axes are normalized as percentages relative to the value at the ``$>20$'' threshold.
As the threshold varies, the overall runtime typically reaches lower value at the threshold of 10 or 20.
The CPWL results show circuit-dependent variability.

\section{Conclusion} 
\label{sec:conclu}

This paper presents a fully differentiable routing formulation for the AMD-Xilinx UltraScale+ FPGA architecture and realizes it in \textit{DiffRouter}, a GPU-accelerated routing framework that optimizes routing demand in continuous space and then legalizes the solution onto the discrete routing resource graph. 
Experiments on the FPGA 2024 Routing Contest benchmarks demonstrate that \textit{DiffRouter} improves runtime and achieves competitive critical-path wirelength over state-of-the-art parallel FPGA routers.
Looking forward, we plan to extend \textit{DiffRouter} to timing-driven routing by incorporating differentiable delay models and timing-aware objectives. 

\section*{Correction Notice}
\label{sec:correction}

This version corrects the total-wirelength (tot WL) results of the paper published in the 2026 IEEE Symposium on Field-Programmable Custom Computing Machines (FCCM), DOI: 10.1109/FCCM68464.2026.00029.
A correction has also been posted on the FCCM 2026 website.
arXiv v1 is identical in content to the published version.

\textbf{Cause of the error.}
Tot WL is an auxiliary metric reported in addition to the two official contest metrics, runtime and CPWL; the FPGA 2024 routing contest does not provide an official tot WL analyzer.
Tot WL was computed by our calculator derived from the contest CPWL analyzer.
In the published implementation, certain local/non-INT routing resources that do not belong to the INT wire types were incorrectly assigned a unit wirelength of 1; under the wirelength convention of the contest, they contribute 0.
In addition, the published Vivado tot WL values were obtained directly from Vivado design checkpoints, whereas the other routers were evaluated from FPGA Interchange Physical Netlists.
In the corrected evaluation, routed Vivado designs are first converted to FPGA Interchange Physical Netlists and evaluated by the same corrected calculator as all other routers.

\textbf{Changes in this version.}
(1)~The tot WL columns of Table~\ref{tab:exp_res} are replaced by the recomputed values, and their geometric-mean ratios are updated (Vivado: 0.870 over the 25 designs it routes; RWRoute: 1.012; CUFR: 1.010; Potter: 1.127; \textit{DiffRouter}: 1.000).
(2)~Section~\ref{sec:exp}-A now states that tot WL is not an official contest metric.
(3)~The published statement that \textit{DiffRouter} ``attains the best CPWL and tot WL among all baselines'' is corrected: \textit{DiffRouter} attains the best CPWL among all baselines and the lowest tot WL among the open-source routers, while Vivado achieves lower tot WL.

\textbf{Unchanged.}
The \textit{DiffRouter} algorithm, the runtime and CPWL results, the ablation studies, and the main conclusions of the paper are unchanged.

\section*{acknowledgement}

This work was supported in part by NSF CCF-2112665 (NSF AI Institute TILOS), AMD, NVIDIA, and the UT iMAGiNE Consortium.

\clearpage
\balance
{
\footnotesize
\bibliographystyle{IEEEtran}
\bibliography{./ref/Top_sim, ./ref/math, ./ref/fpga}
}

\end{document}